\documentclass[conference]{IEEEtran}
\IEEEoverridecommandlockouts
\usepackage{cite}
\usepackage{amsmath,amssymb,amsfonts}
\usepackage{algorithmic}
\usepackage{graphicx}
\usepackage{textcomp}
\usepackage{xcolor}
\usepackage{lineno}
\usepackage{float}
\usepackage{multirow}
\usepackage{CJKutf8}
\usepackage[hidelinks]{hyperref}
\usepackage{subcaption}
\usepackage{listings}

\def\BibTeX{{\rm B\kern-.05em{\sc i\kern-.025em b}\kern-.08em
    T\kern-.1667em\lower.7ex\hbox{E}\kern-.125emX}}
\begin{document}

\title{H4M: Heterogeneous, Multi-source, Multi-modal, Multi-view and Multi-distributional Dataset for Socioeconomic Analytics in the Case of Beijing
\thanks{\textsuperscript{*}Corresponding author.\\
\indent This work is supported in part by The National Natural Science Foundation of China (Grant No. 71871006), in part by ACCESS –- AI Chip Center for Emerging Smart Systems, Hong Kong SAR, Hong Kong Research Grant Council (Grant No. 27206321), and National Natural Science Foundation of China (Grant No. 62122004).}
}

\makeatletter
\newcommand{\linebreakand}{%
  \end{@IEEEauthorhalign}
  \hfill\mbox{}\par
  \mbox{}\hfill\begin{@IEEEauthorhalign}
}
\makeatother

\author{\IEEEauthorblockN{Yaping Zhao\textsuperscript{1,2}}
\IEEEauthorblockA{
zhaoyp@connect.hku.hk}
\and
\IEEEauthorblockN{Shuhui Shi\textsuperscript{1}}
\IEEEauthorblockA{
shishuhui.hit@gmail.com}
\and
\IEEEauthorblockN{Ramgopal Ravi\textsuperscript{1}}
\IEEEauthorblockA{
raviramgopal@gmail.com}
\linebreakand
\IEEEauthorblockN{Zhongrui Wang\textsuperscript{1,2}}
\IEEEauthorblockA{
zrwang@eee.hku.hk}
\and
\IEEEauthorblockN{Edmund Y. Lam\textsuperscript{1,2}}
\IEEEauthorblockA{
elam@eee.hku.hk}
\and
\IEEEauthorblockN{Jichang Zhao\textsuperscript{3,*}}
\IEEEauthorblockA{
jichang@buaa.edu.cn}
\linebreakand
\IEEEauthorblockN{\textsuperscript{1}\textit{The University of Hong Kong}}
\and
\IEEEauthorblockN{\textsuperscript{2}\textit{ACCESS–-AI Chip Center for Emerging Smart Systems}}
\and
\IEEEauthorblockN{\textsuperscript{3}\textit{Beihang University}}
}

\maketitle

\begin{abstract}
The study of socioeconomic status has been reformed by the availability of digital records containing data on real estate, points of interest, traffic and social media trends such as micro-blogging. In this paper, we describe a heterogeneous, multi-source, multi-modal, multi-view and multi-distributional dataset named  \textit{``H4M"}. The mixed dataset contains data on real estate transactions, points of interest, traffic patterns and micro-blogging trends from Beijing, China. The unique composition of H4M makes it an ideal test bed for methodologies and approaches aimed at studying and solving problems related to real estate, traffic, urban mobility planning, social sentiment analysis etc. The dataset is available at: \href{https://indigopurple.github.io/H4M/index.html}{\textcolor{blue}{https://indigopurple.github.io/H4M/index.html}}.
\end{abstract}


\section*{Background \& Summary}

The availability of extensive data has helped provide fundamental and qualitative insights on socioeconomic analytics. This is down to two main factors - global adoption of digital records and the growing use of e-services. As Figure~\ref{fig:teaser} shows, we propose a heterogeneous,
multi-source, multi-modal, multi-view and multi-distributional
dataset named “H4M”. The mixed dataset contains data on real
estate transactions, points of interest, traffic patterns and micro-
blogging trends from Beijing, China.

Firstly, the vast amount of data obtained from real estate transaction records can effectively reflect socioeconomic status. Although a previous work~\cite{london,fairfax,ames} helped collect and establish a model for estimation and prediction, the datasets used were either outdated or limited. Due to the rapid expansion and growth of real estate in China, we have been able to collect a large amount of data as shown in Figure~\ref{fig:scatter_rs} and Table~\ref{tab:comp_house}.

\begin{figure}
    \centering
    \vspace{-10mm}
    \includegraphics[width=\linewidth]{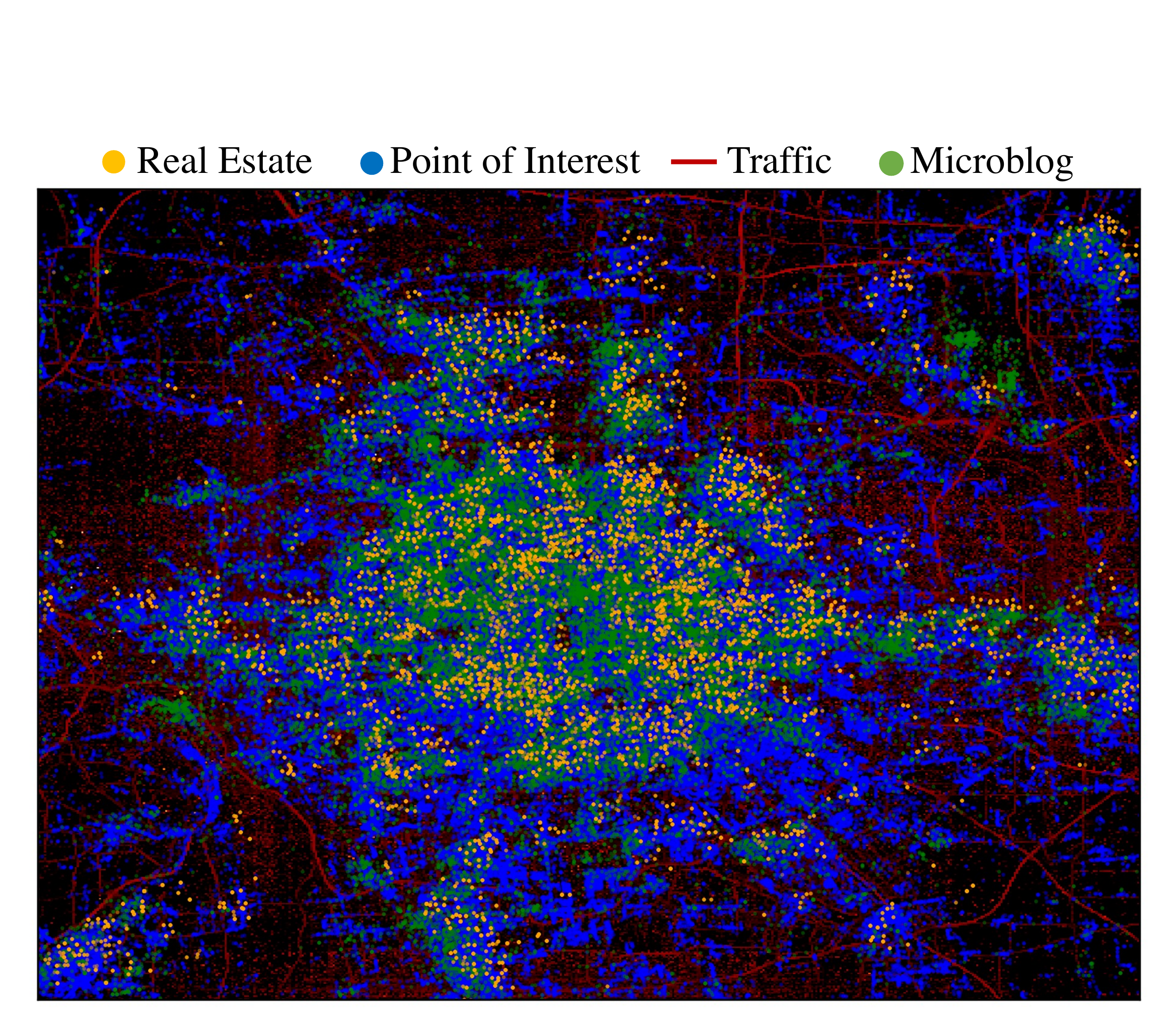}
    \caption{The \textit{``H4M"} dataset contains data on real estate transactions, points of interest, traffic patterns and micro-blogging trends from Beijing, China.}
    \label{fig:teaser}
\end{figure}

\begin{figure*}
  \begin{subfigure}{0.5\linewidth}
    \centering\includegraphics[width=\linewidth, trim=0 100 0 0,clip]{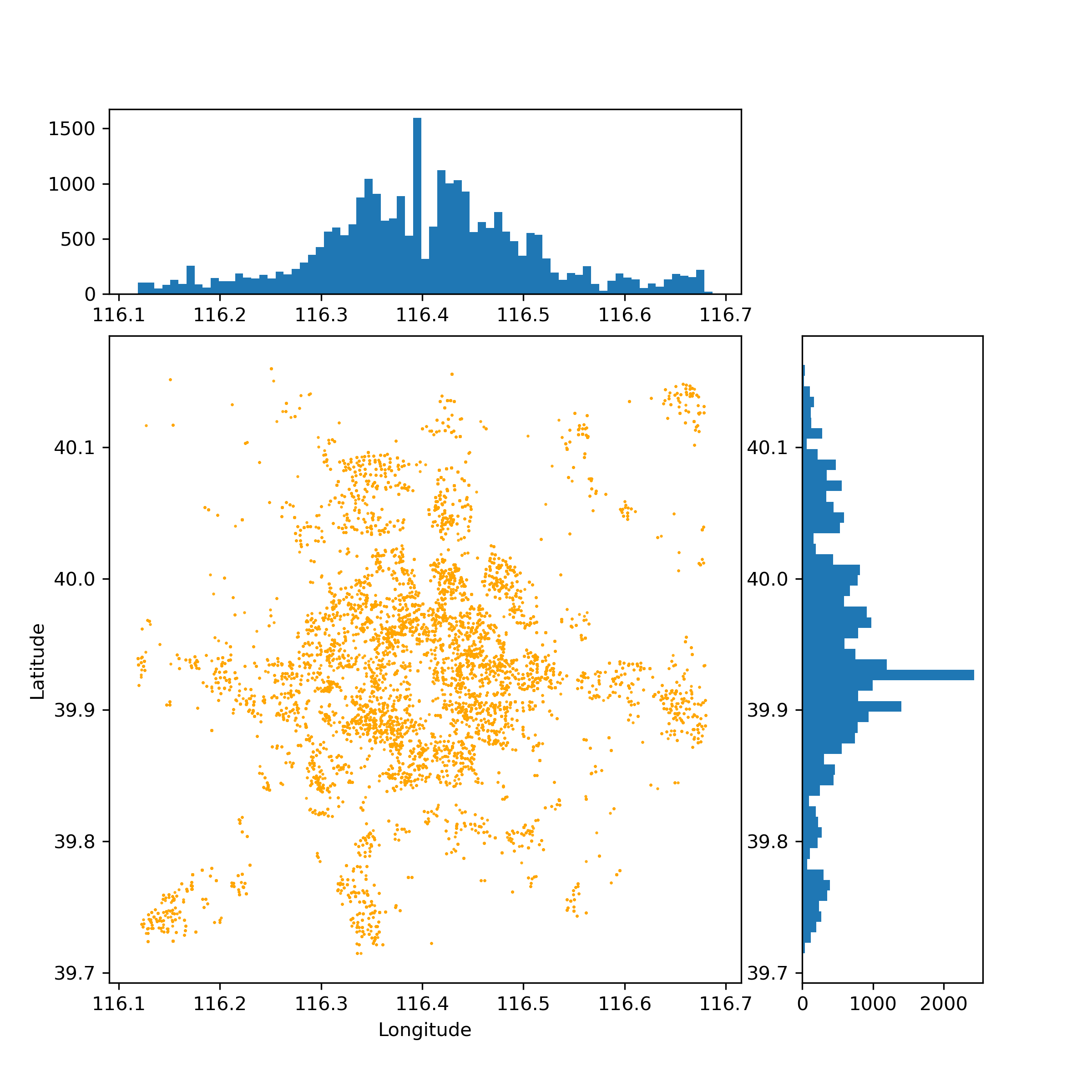}
    \caption{Real Estate}
    \label{fig:scatter_rs}
  \end{subfigure}
  \begin{subfigure}{0.5\linewidth}
    \centering\includegraphics[width=\linewidth, trim=0 100 0 0,clip]{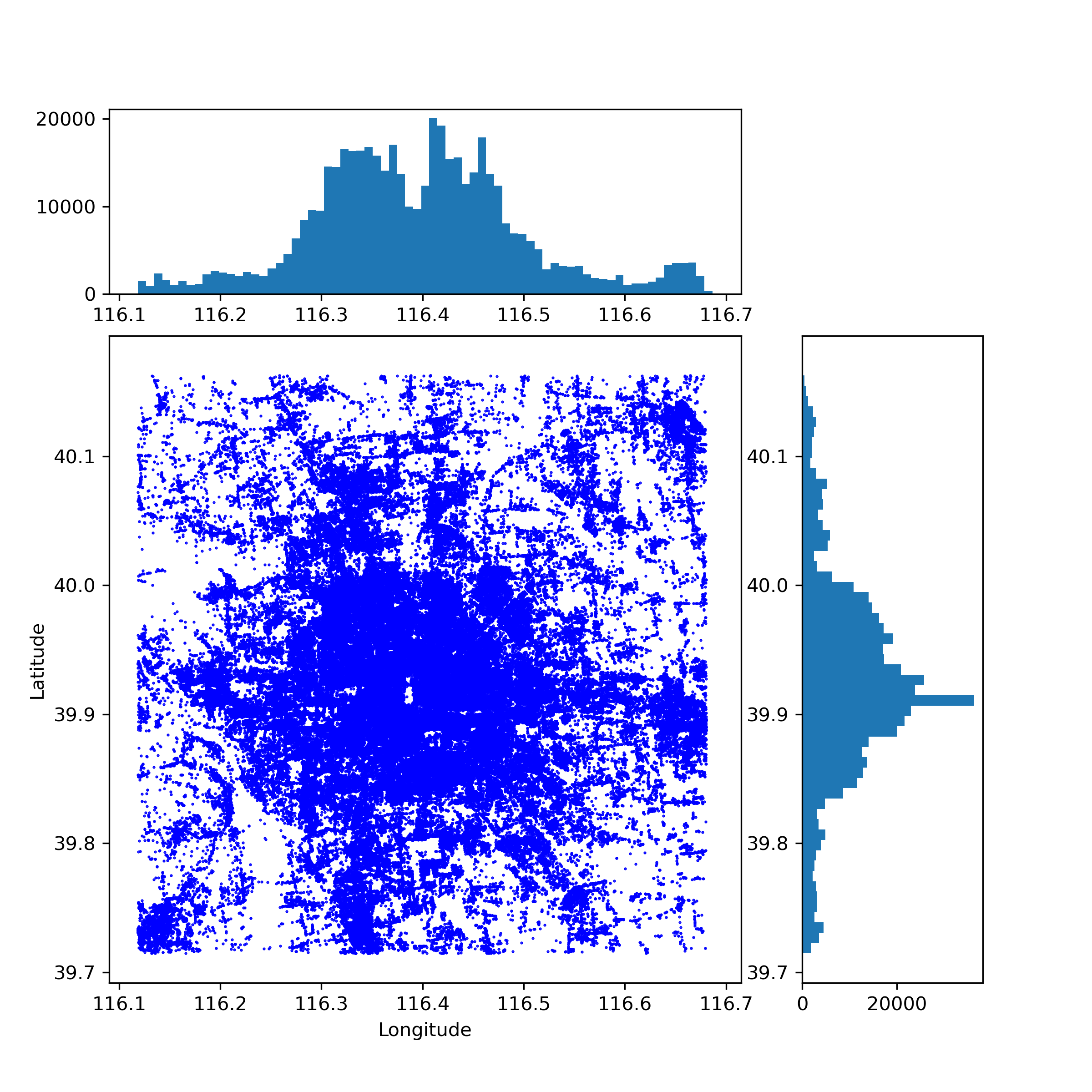}
    \caption{Points of Interest}
    \label{fig:scatter_poi}
  \end{subfigure}
  \begin{subfigure}{0.5\linewidth}
    \centering
    \vspace{10mm}
    \includegraphics[width=\linewidth, trim=30 0 50 0,clip]{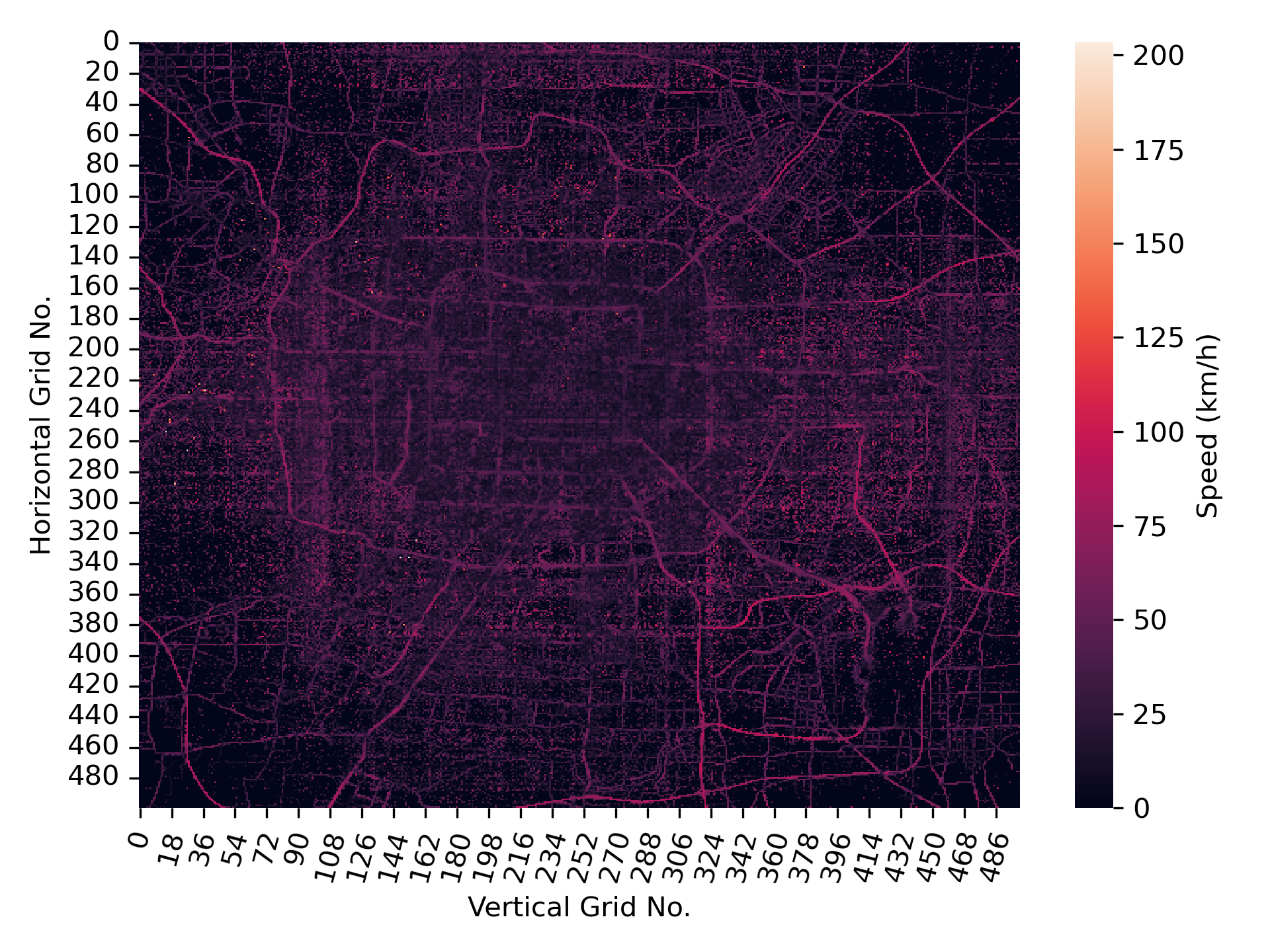}
    \caption{Traffic}
    \label{fig:scatter_traffic}
  \end{subfigure}
  \begin{subfigure}{0.5\linewidth}
    \centering
    \includegraphics[width=\linewidth, trim=0 100 0 150,clip]{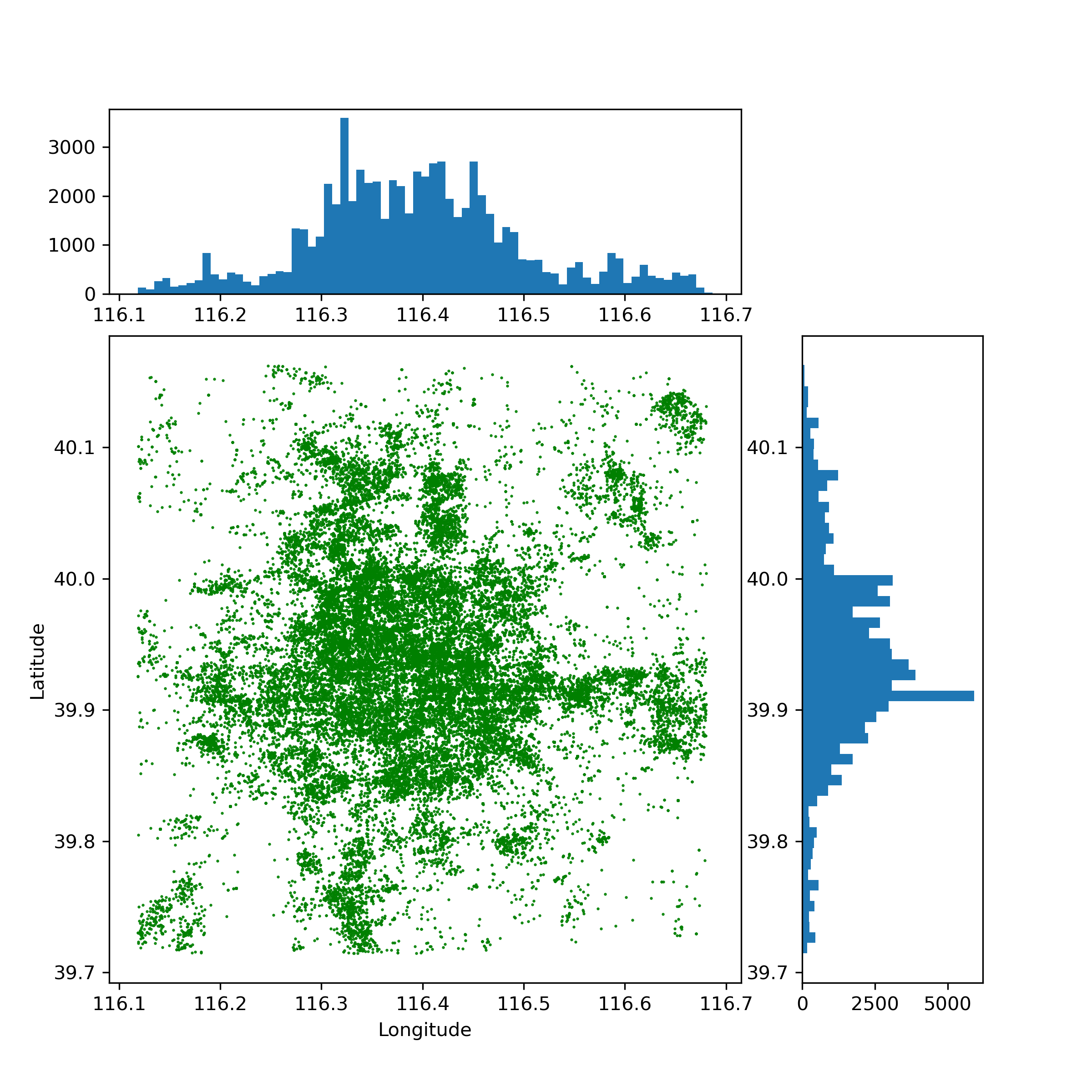}
    \caption{Microblog}
    \label{fig:scatter_weibo}
  \end{subfigure}
  \caption{The scatter plot with histograms of (a) real estate, (b) points of interest, (d) microblog posts, and the heat map of (c) traffic patterns. The distributions of real estate, points of interest and microblog data present complicated patterns and are distinguished from each other. Due to the data collection and processing method, the traffic data is uniformly distributed. Therefore we visualize the average traffic speed of a whole day in each grid.}
  \label{fig:scatter}
\end{figure*}

\begin{table*}
\centering
\begin{tabular}{|l|l|l|l|l|}
\hline
Dataset &  Lodon~\cite{london}& Fairfax~\cite{fairfax} & Ames~\cite{ames} & H4M (Ours) \\
\hline
Data    &  2,108 & 15,135  & 2,930   & 28,645  \\
\hline
Year    &  2001  & 2004 - 2007  & 2006 - 2010 & 2017 - 2018 \\
\hline
Area & London, UK & Fairfax, Virginia, USA  & Ames, Iowa, USA & Beijing, China \\
\hline
\end{tabular}
\vspace{2mm}
\caption{\label{tab:comp_house}Comparisons of different real estate datasets described in terms of amount of data, recorded year and geographical area.}
\end{table*}

\begin{table*}
\centering
\begin{tabular}{|l|l|l|}
\hline
Dataset & Lagos~\cite{lagos} & H4M (Ours) \\
\hline
Data  & 157  & 497,256  \\
\hline
Types & 9 & 18 \\ 
\hline
Area & Lagos, Nigeria & Beijing, China \\
\hline
\end{tabular}
\vspace{2mm}
\caption{\label{tab:comp_infra}Comparisons of different points of interest described in terms of data amount, type amount and geographical area.}
\end{table*}

\begin{table*}
\centering
\begin{tabular}{|l|l|l|l|}
\hline
Dataset & PeMS~\cite{pems} & Hangzhou~\cite{tian2018lstm} & H4M (Ours) \\
\hline
Data  & 39,000  & 34 & 23,772  \\
\hline
Year & 2019 & 2015-2016 & 2017 \\ 
\hline
Area & California, USA & Hangzhou, China & Beijing, China \\
\hline
\end{tabular}
\vspace{2mm}
\caption{\label{tab:comp_traffic}Comparisons of different traffic datasets described in terms of data amount, recorded year and geographical area.}
\end{table*}

\begin{table*}
\centering
\begin{tabular}{|l|l|l|l|l|}
\hline
Dataset & Short-Text~\cite{short_text} & WeiboRank~\cite{weibo_rank} & Weibo-COV~\cite{weibo_cov}  & H4M (Ours) \\
\hline
Data  & 4,6345  & 22,620,281 & over 40 million & over 100 million  \\
\hline
Geolocated & no  & no & no & yes \\ 
\hline
\end{tabular}
\vspace{2mm}
\caption{\label{tab:comp_microblog}Comparisons of different microblog datasets described in terms of data amount and whether its geolocated or not.}
\end{table*}

\begin{table*}[h]
\centering
\begin{tabular}{|l|l|l|l|l|}
\hline
Data type & House & Points of Interest & Traffic flow  & Microblog \\
\hline
Issuer  & Lianjia  & Baidu Maps & Baidu Maps & Sina Weibo  \\
\hline
Acquisition & Web crawler  & Web crawler & Web crawler & Web crawler \\ 
\hline
Format & JSON & JSON & Pickle & Text\\
\hline
\end{tabular}
\vspace{2mm}
\caption{\label{tab:method}An overview of the different data sources and
the methods used to process them.}
\end{table*}

Secondly, the availability of points of interest data has helped define a novel area of research - We are now able to model urban structures and predict socioeconomic indicators~\cite{lagos}. Compared to previous work in this field, our dataset offers more diverse data as shown in Figure~\ref{fig:scatter_poi} and Table~\ref{tab:comp_infra}. 

The Caltrans Performance Measurement System (PeMS)~\cite{pems} is the most extensively used dataset in traffic flow prediction - the data is collected from California, USA and the existing dataset~\cite{tian2018lstm} from China is insufficient. For this reason, we provide traffic flow data with amount in parallel to PeMS, as shown in Figure~\ref{fig:scatter_traffic} and Table~\ref{tab:comp_traffic}. 

Finally the emergence of geolocated information and social media services like microblogs has provided us opportunities to quantitatively inspect social wellness~\cite{quercia2012tracking} and also the socioeconomic status of geographical regions~\cite{llorente2015social}. Since the existing datasets~\cite{short_text,weibo_rank,weibo_cov} did not have the geographic coordinates associated with its posts, we could not geolocate each data point. To overcome this, we collected over 100 million Chinese microblog posts with location coordinates as show in Figure~\ref{fig:scatter_weibo} and Table~\ref{tab:comp_microblog}.

It is important to note that the aforementioned datasets from previous work only offer data related to a single aspect for socioeconomic analytics. Diverse data should be leveraged to guarantee comprehensive analysis. Barlacchi \textit{et. al.}~\cite{milan} proposed a multi-source dataset on two geographical areas - the city of Milan and the province of Trentino. This dataset is a multi-source aggregation of telecommunications, weather, news, social network and electricity data. In contrast, H4M comprises of data on real estate, points of interest, traffic and microblogs with geographical coordinates from Beijing, China. Therefore, it intrinsically owns some complex characteristics such as heterogeneous, multi-source, multi-modal, multi-view and multi-distributional.

\subsection*{Heterogeneous}
More specifically, H4M contains $28,645$ entries of real estate data, $497, 256$ pieces of points of interest, $250, 000$ grids of traffic data and over 100 million of microblog posts in Beijing. These are shown in Figure~\ref{fig:scatter} as scatter plots and a heat map. From this we can conclude that the H4M dataset is heterogeneous.

\subsection*{Multi-source}
As seen in Table~\ref{tab:method}, by using web crawlers, we access and collect data by scraping a large number of websites. Given that H4M contains data from different souces, we can assume it to be a multi-source dataset.

\subsection*{Multi-modal}
As Table~\ref{tab:method} shows, each type of data is stored in a specific format. The data formats used in H4M include JSON, Pickle and plain text thus making H4M multi-modal.

\subsection*{Multi-view}
The unique composition of H4M makes it an ideal test bed for methodologies and approaches aimed at studying and solving problems related to diverse and different perspectives including real estate, traffic, urban mobility planning, social sentiment analysis etc. Therefore, the H4M dataset is multi-view.

\subsection*{Multi-distributional}
The distributions of real estate, points of interest and microblog data are respectively shown in Figure~\ref{fig:scatter_rs},~\ref{fig:scatter_poi},~\ref{fig:scatter_weibo}. This represents complicated patterns and are distinguished from each other. Due to the data collection and processing method that is used, the traffic data is uniformly distributed. Therefore, the H4M dataset is multi-distributional.

The intrinsic heterogeneous, multi-source, multi-modal, multi-view and multi-distributional characteristics of H4M make it important and novel for socioeconomic analytics. For instance, the aforementioned characteristics can be an invaluable resource to assess the economic and social value of real estate and thus boost the accuracy of real estate valuation modeling by estimating many factors of property~\cite{zhao2022pate}. Moreover, it allows for predicting large-scale traffic congestion based on multi-modal fusion and representation mapping~\cite{zhou2022traffic}.

To conclude, we believe that the \textit{``H4M"} dataset will aid and encourage researchers to design algorithms capable of exploiting various socioeconomic indicators.

\begin{figure*}[h]
\centering
\includegraphics[width=0.8\linewidth]{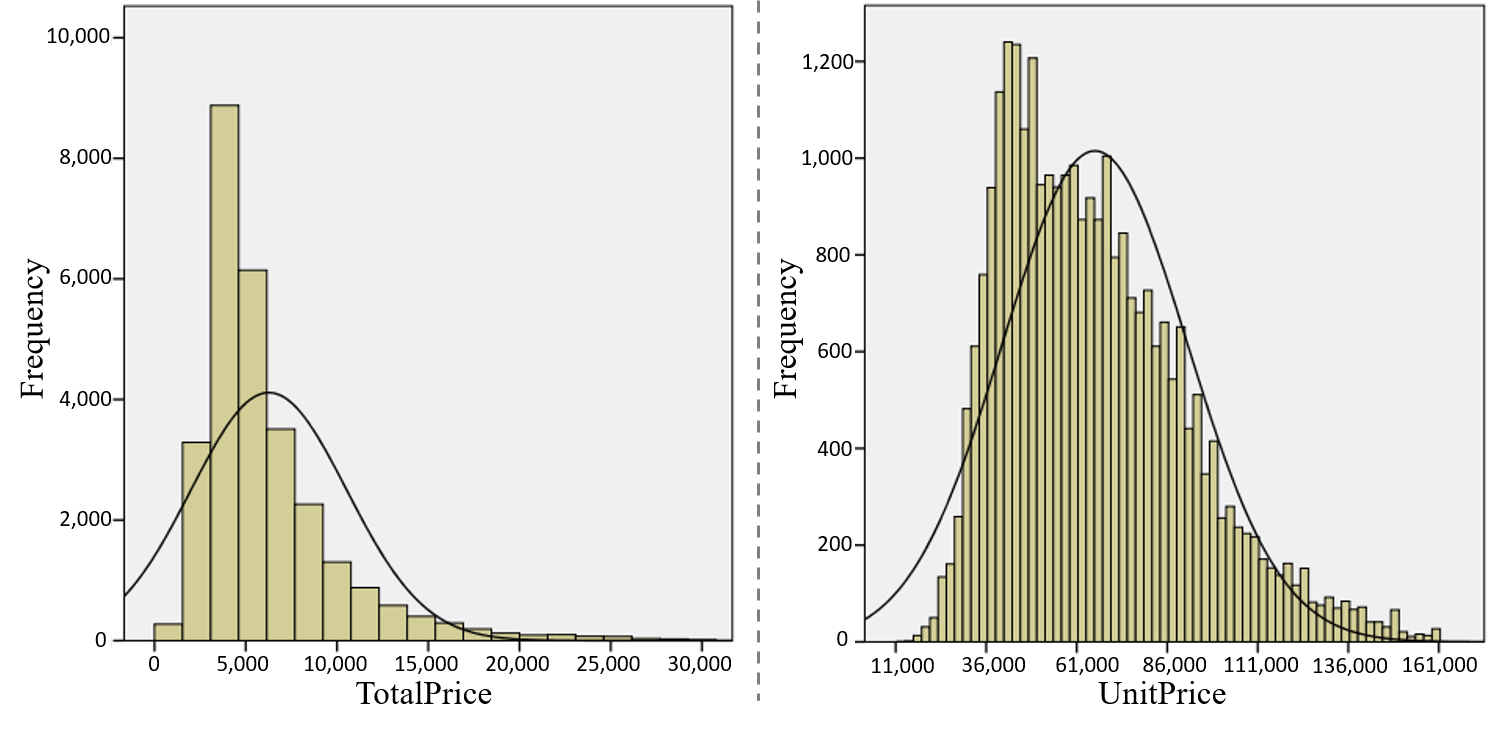}
\caption{The distribution of total price and unit price per square meter of houses in Beijing by histogram with distribution curve.}
\label{fig:house_price}
\end{figure*}

\section*{Methods}
In this section, we describe the collection and processing methods of our multi-source dataset. After processing, we aggregate all data and associate it with geographical coordinates. This allows for easier comparisons between different areas thus reducing the overload of geographical management.  Table~\ref{tab:method} provides an overview of the different data sources and the methods used to process them.

\subsection*{Real Estate}
We use a web crawler to browse through  Lianjia ‘s website (\href{https://bj.lianjia.com/}{https://bj.lianjia.com/}). The crawler indexes numerous pages on real estate transactions and retrieves them as HTML files. We then decompose the files and identify the relevant transaction data after which they are extracted and collected.

The data from Lianjia contains real estate addresses which is used to send requests to Baidu Maps (\href{https://map.baidu.com/}{https://map.baidu.com/}) to obtain geographical coordinates (latitude and longitude) for each housing record. Finally the transaction data is organized and stored as JSON files.

\subsection*{Points of Interest}
Using Baidu Maps, we identify and collect points of interest (POI) and their geographical coordinates in Beijing. We then enumerate the POIs and check their points of interest type. The POIs with meaningless types are discarded. The remaining 18 types of points of interest are stored as JSON files.

\subsection*{Traffic}
Our traffic flow data is collected from Baidu Maps. The latitudinal and longitudinal values for Beijing and its different regions vary between range $116.1186218\sim116.6802978$, meanwhile latitude values vary in $39.7145817\sim40.1626081$. For convenience, we evenly divide the regions in the above range into $500 \times 500 = 250,000$ grids, as shown in Figure~\ref{fig:grids}. Grids without roads or insufficient traffic data are padded with zero values, 
and for the remaining 23772 grids, we collect traffic average speed information per $5$ minutes from 6 \textit{a.m} to \textit{12 a.m}. These are then stored in Pickle files.

\begin{figure}[h]
\centering
\includegraphics[width=0.65\linewidth]{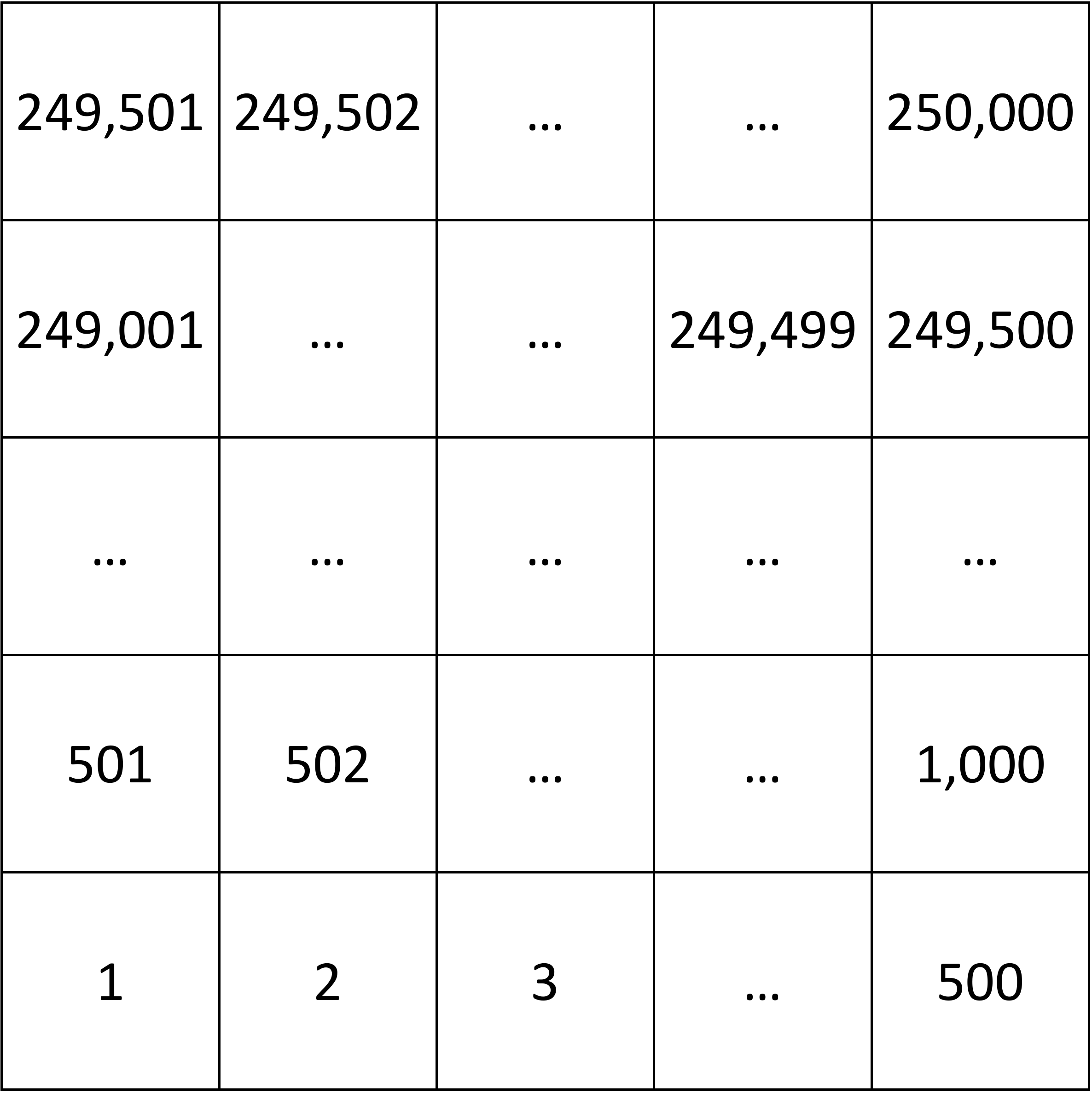}
\caption{The grid system employed in traffic data.}
\label{fig:grids}
\end{figure}

\subsection*{Microblog Posts}
We use a web crawler to access the website of Sina Weibo (\href{https://weibo.com/}{https://weibo.com/}); the most commonly used microblog platform in China.  We then identify microblog posts with location information and discard the remaining data points. Finally, over $100$ millions microblog posts between September 12 2013 and April 20 2015 are collected and stored as text files. 

\begin{figure*}[h]
\centering
\includegraphics[width=0.7\linewidth]{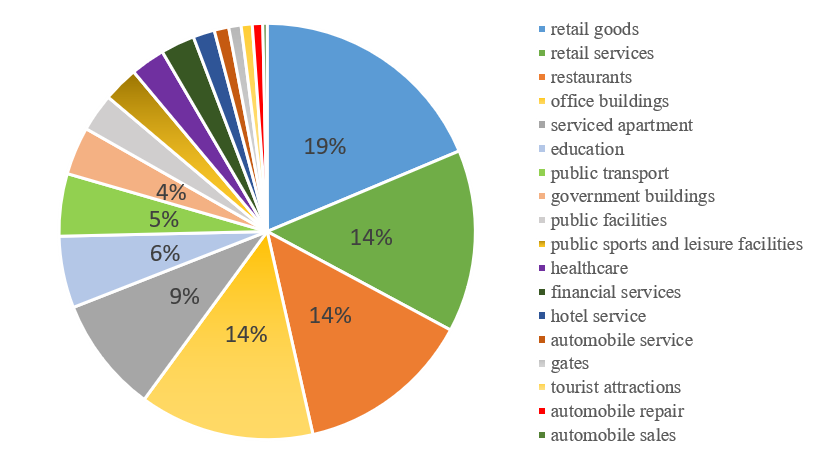}
\caption{Pie chart containing the percentages of different points of interest types in Beijing.}
\label{fig:infra_pct}
\end{figure*}

\begin{figure}[h]
\centering
\includegraphics[width=\linewidth]{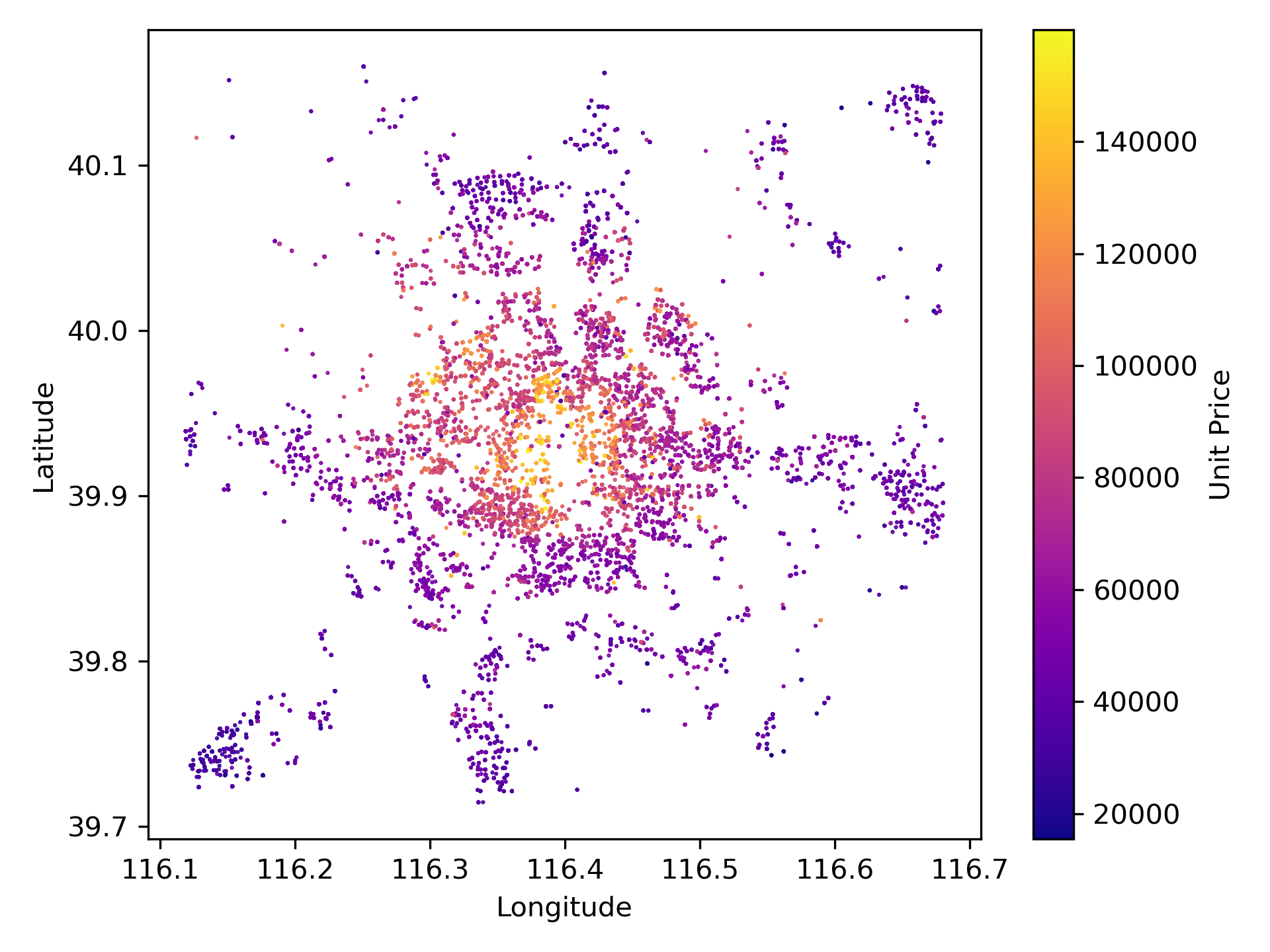}
\caption{The scatter plot of real estate in Beijing, where the colors of the scatter points vary according to the unit price per square
meter of house.}
\label{fig:house_dist}
\end{figure}

\begin{figure*}[h]
\centering
\includegraphics[width=0.7\linewidth]{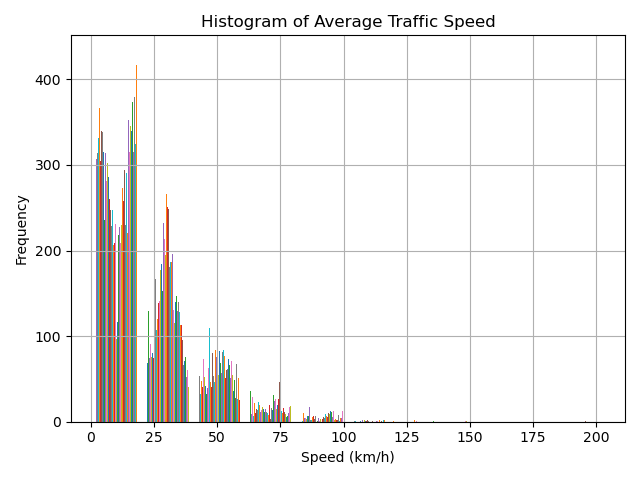}
\caption{The histogram of average traffic speed of a whole day in each grid of Beijing, where the grid system employed is given in Figure~\ref{fig:grids}.}
\label{fig:hist_traffic}
\end{figure*}

\begin{figure*}[h]
  \begin{subfigure}{0.5\linewidth}
    \centering\includegraphics[width=\linewidth, trim=0 50 0 0,clip
    ]{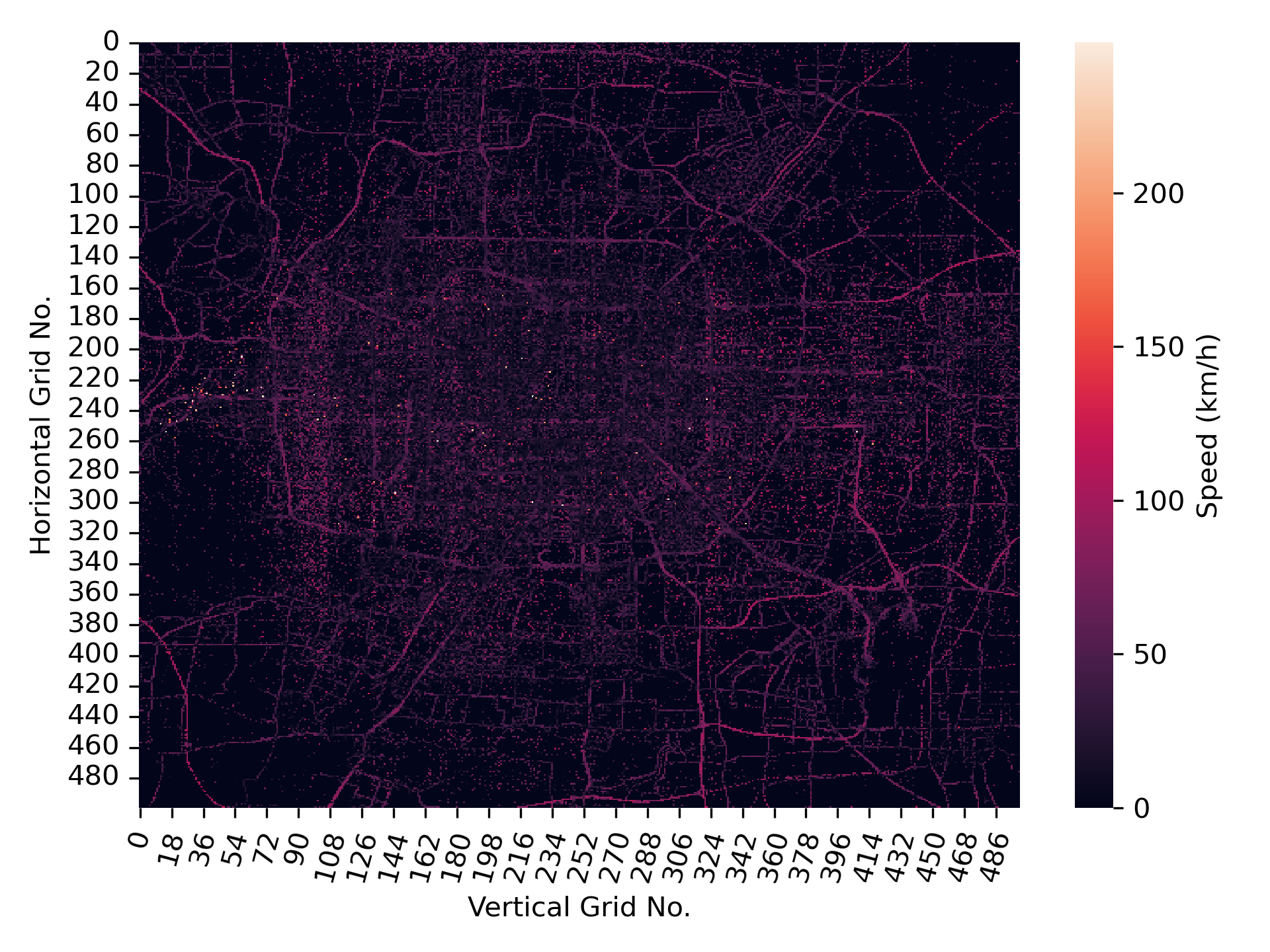}
    \caption{The heat map of traffic speed at 8 am.}
  \end{subfigure}
  \begin{subfigure}{0.5\linewidth}
    \centering\includegraphics[width=\linewidth, trim=0 50 0 0,clip]{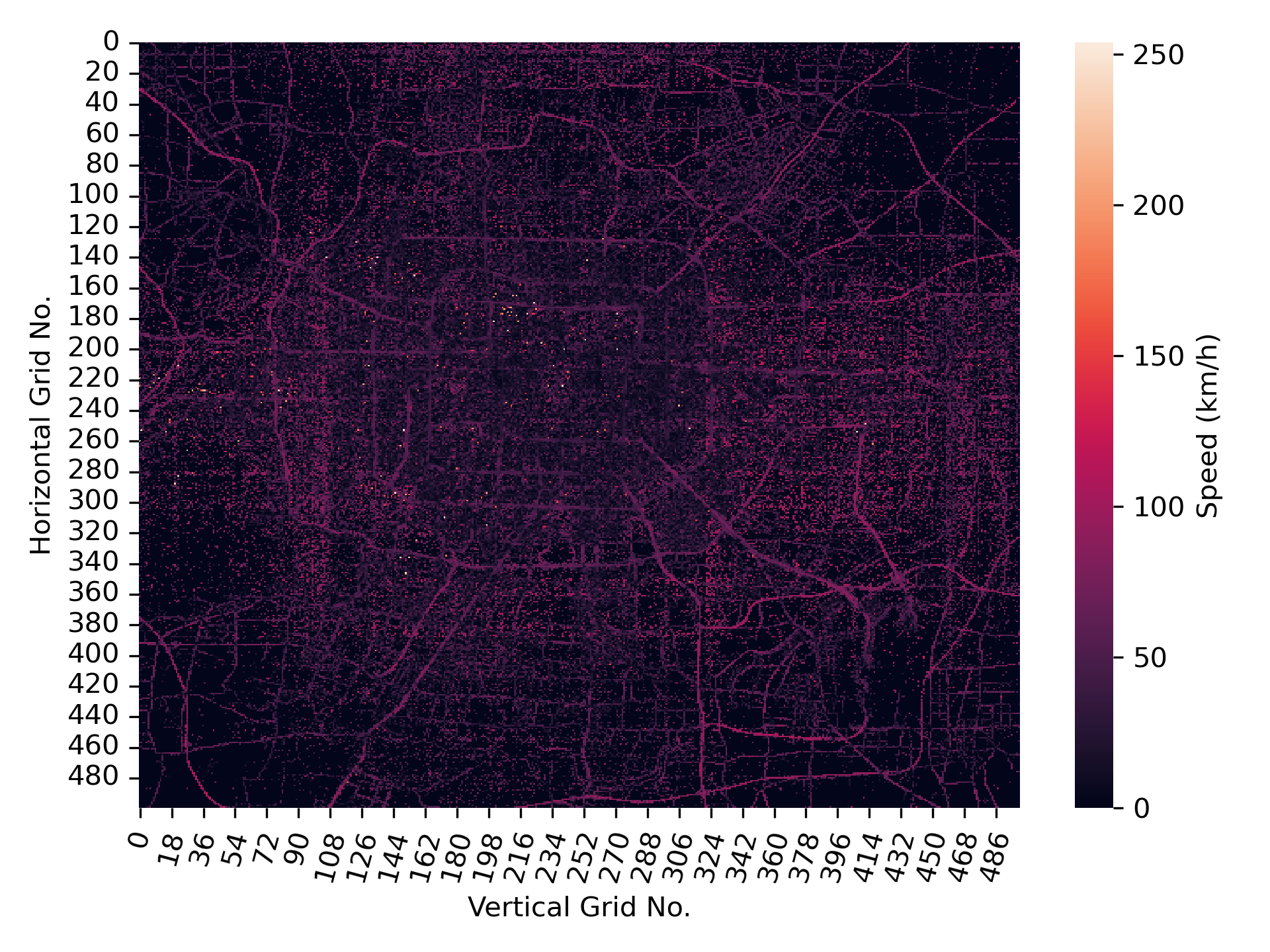}
    \caption{The heat map of traffic speed at 12 am.}
  \end{subfigure}
  \begin{subfigure}{0.5\linewidth}
    \centering\includegraphics[width=\linewidth, trim=0 50 0 0,clip]{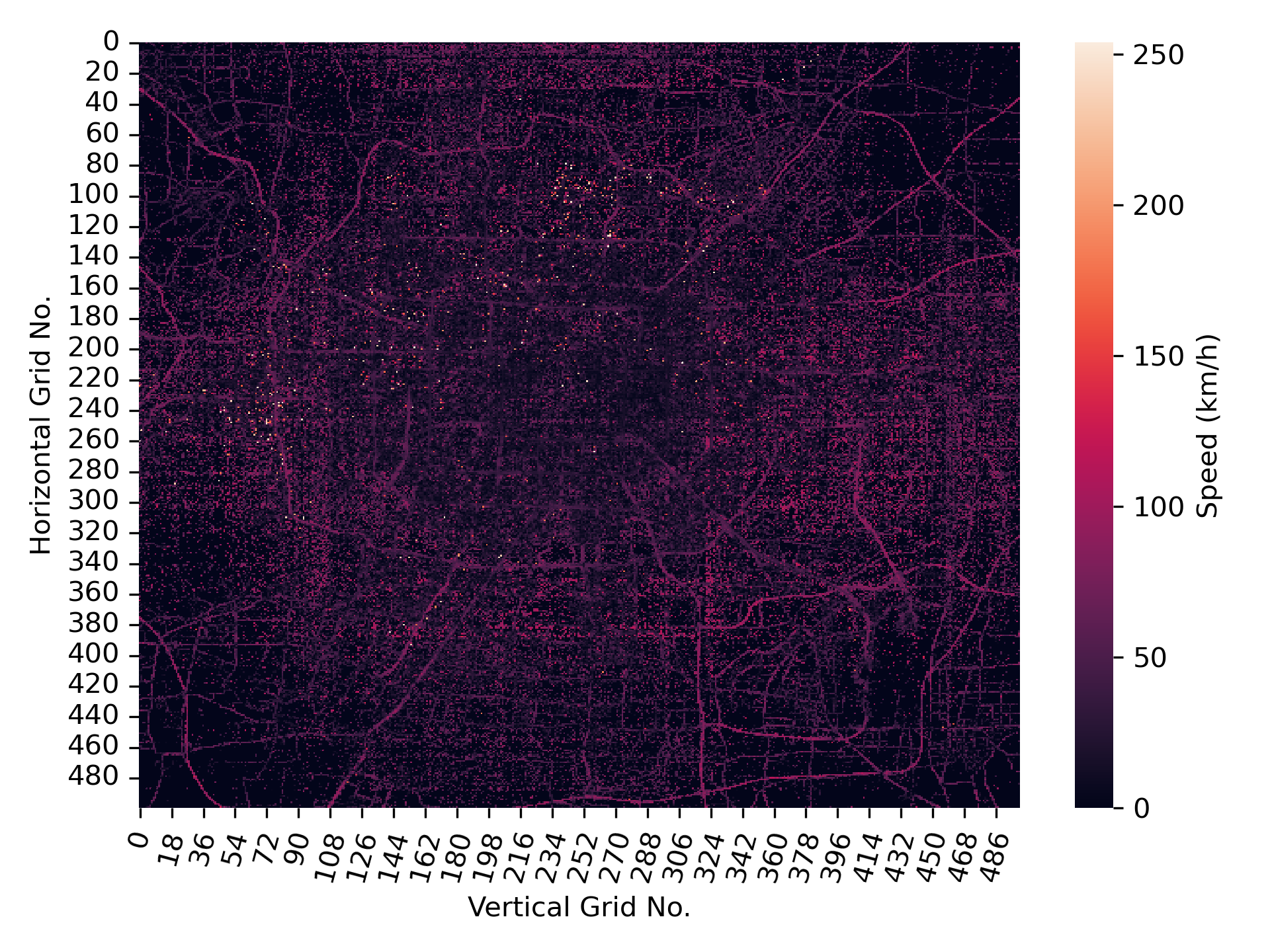}
    \caption{The heat map of traffic speed at 4 pm.}
  \end{subfigure}
  \begin{subfigure}{0.5\linewidth}
    \centering\includegraphics[width=\linewidth, trim=0 50 0 0,clip]{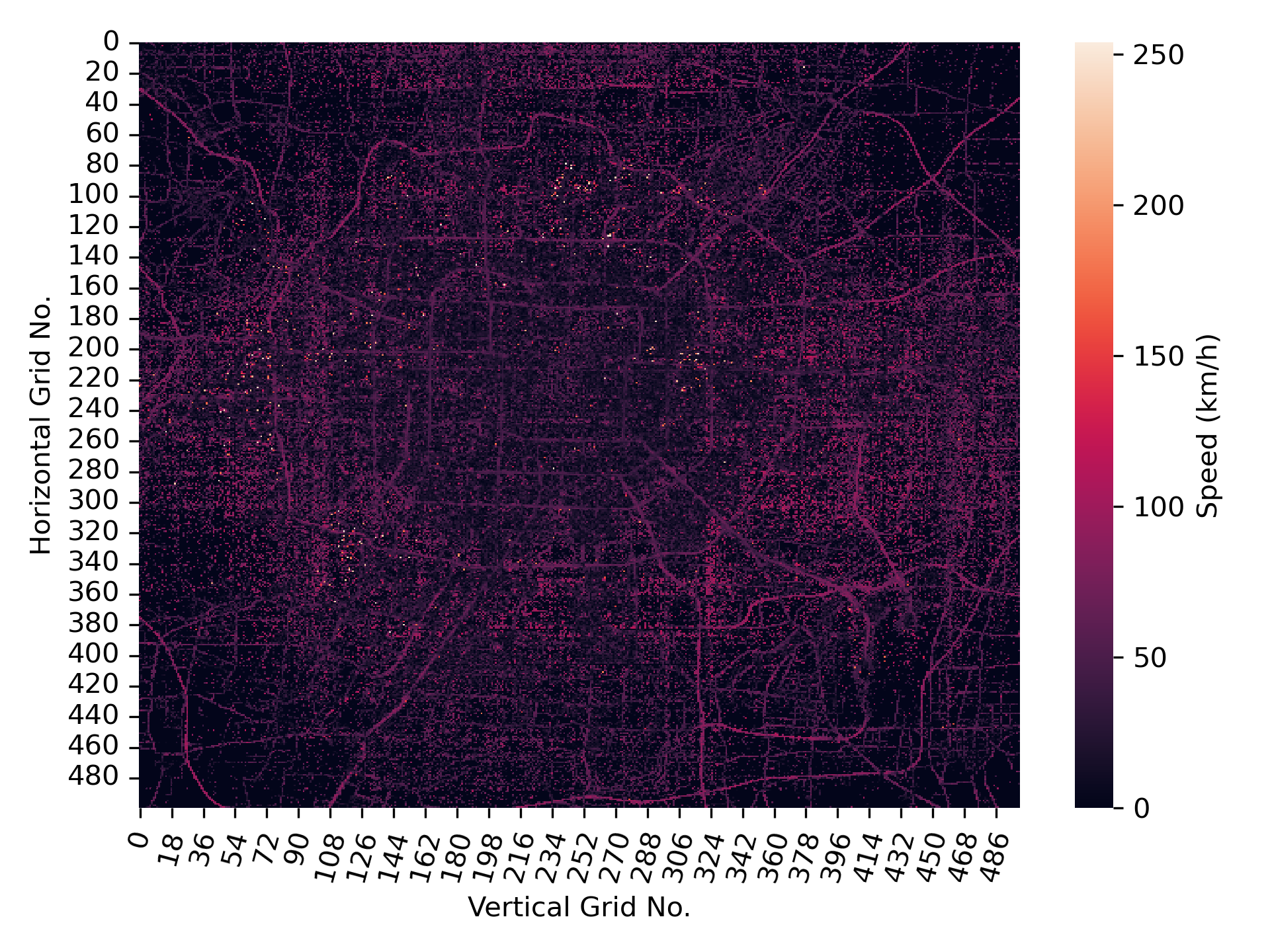}
    \caption{The heat map of traffic speed at 8 pm.}
  \end{subfigure}
  \caption{The heat map of traffic speed in Beijing at (a) 8 am, (b) 12 pm, (c) 4 pm, (d) 8 pm, respectively, where the grid system employed is given in Figure~\ref{fig:grids}.}
  \label{fig:traffic_day}
\end{figure*}

\begin{table}[h]
\centering
\begin{tabular}{|l|l|l|}
\hline
\multirow{ 2}{*}{Variable} & TotalPrice & UnitPrice \\
& (thousand RMB) & (in RMB)\\
\hline
Avg & 6,264  &  66,008 \\
\hline
Med & 5,000  & 61,880 \\ 
\hline
Std & 4,272  & 25,587 \\ 
\hline
Min & 600 & 13,209\\
\hline
Max & 95,000 & 159,975\\
\hline
Pct25 & 3,750 & 45,694 \\
\hline
Pct50 & 5,000 & 61,880 \\
\hline
Pct75 & 7,350 & 81,541\\
\hline
\end{tabular}
\vspace{2mm}
\caption{\label{tab:house_stat}House price statistics showing the average, median, standard deviation, minimum, maximum, 25th percentile, 50th percentile and 75th percentile values of house price in Beijing. TotalPrice is the total price of a house in thousand Renmibi (RMB), UnitPrice is the unit price per square meter in RMB. }
\end{table}

\section*{Technical Validation}
An accurate techincal validation of the dataset is limited due to the absence of similar multi-source datasets that are available for comparison. Hence, in this section we propose a statistical and visual characterization with the aim of supporting the naive correctness of the information provided.

\subsection*{Real Estate}
Figure~\ref{fig:house_price} represents the distribution of total price and unit price per square meter of houses in Beijing by histogram with distribution curve. The maximum and minimum values for the total price of house are approximately 6.4 and 1.1 standard deviations away from the mean. The total price distribution has a skewness of 3.24. In comparison, the maximum and minimum values for UnitPrice (price per square meter) approximately 3.7 and 2 standard deviations away from the mean with a skewness of 0.81.

Table~\ref{tab:house_stat} describes the average, standard deviation, minimum, maximum, 25th percentile, 50th percentile and 75th percentile values of house price in Beijing. Those values are reasonable.
Figure~\ref{fig:house_dist} shows the geographical distribution and unit price of real estate in Beijing. The houses closer to the city center are more expensive, while the houses in the suburbs far away from the city are rarer and cheaper.

\subsection*{Points of Interest}
Figure~\ref{fig:infra_pct} is a pie chart containing the percentages of different points of interest types in Beijing. Going clockwise, Retail goods and services contribute 19 percent and 14 percent of all goods and services. Restaurants, commercial/office buildings  and serviced apartments makeup 37 percent of the pie chart. As Beijing’s subway system is the primary mode of public transport in the city, this points of interest type contributes only around 5 percent. All the other points of interest types complete the remaining 25 percent.

\begin{figure*}[h]
\centering
\includegraphics[width=\linewidth]{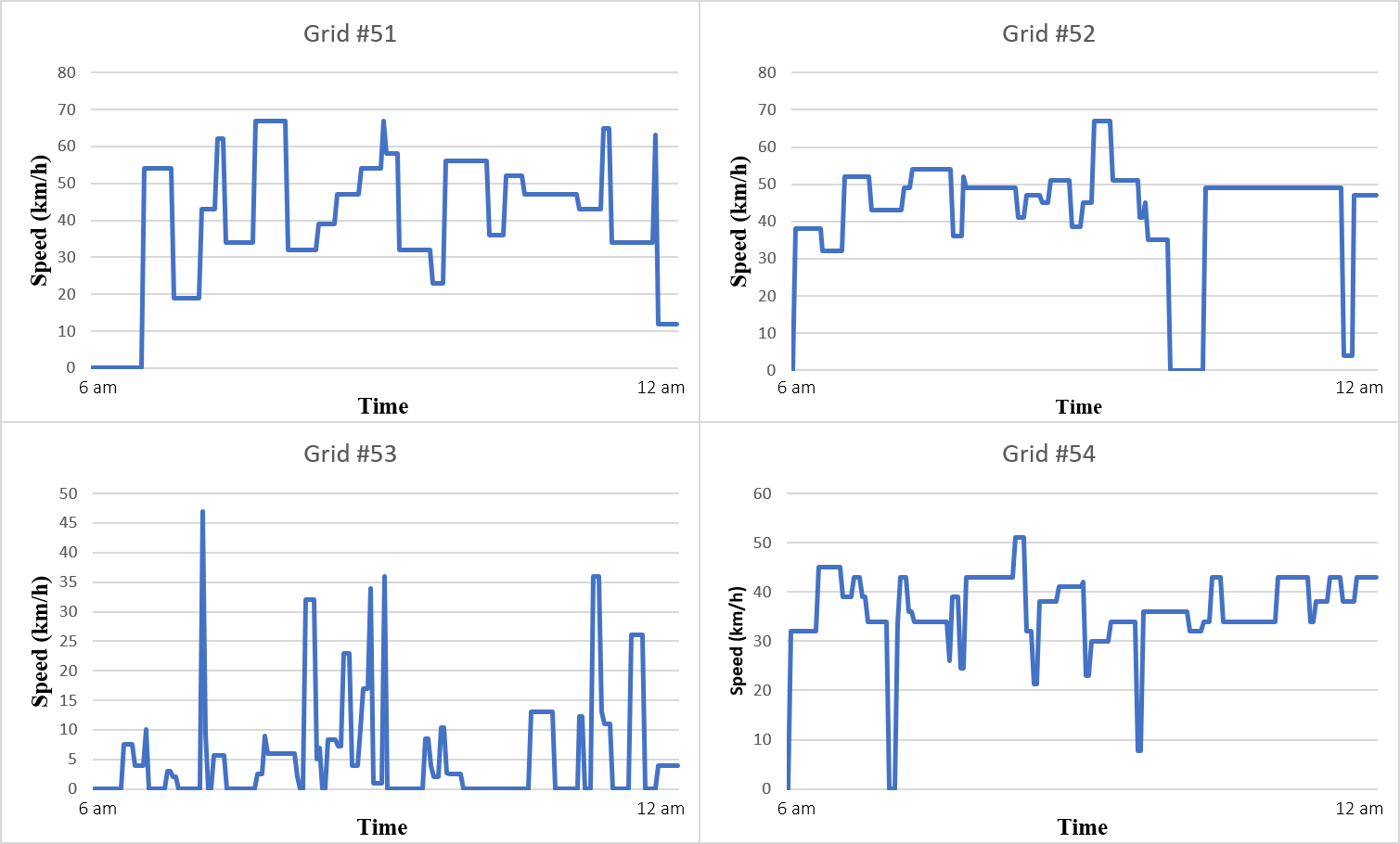}
\caption{Visualization of traffic flow in a day in grids 51-54 in Beijing.}
\label{fig:traffic_data}
\end{figure*}

\subsection*{Traffic}
To testify the validation of our collected traffic data, Figure~\ref{fig:hist_traffic} presents the histogram of average traffic speed of a whole day in each grid of Beijing. We observe that most grids load the traffic flow with average speed as $0\sim50$ km/h, and the average speeds are generally below $100$ km/h (except for very few between $100$ and $120$ km/h). It is reasonable and valid, since generally traffic tends to be concentrated in a few areas; on another hand,
the speed limit on Beijing's roads is generally below $100$ km/h, while the maximum speed limit on airport expressways in Beijing is $120$ km/h.

In addition, we visualize the traffic speed at different time in a day as shown in Figure~\ref{fig:traffic_day}. Moreover, we randomly choose 4 grids and visualize the traffic flow in a day as shown in Figure~\ref{fig:traffic_data}. The figure contains traffic patterns for grids 51-54 in Beijing. The Y-axis in each plot is the average speed of traffic in km/h and the X-axis is time. As mentioned earlier, traffic speeds are obtained using a web crawler from Baidu Maps every 5 minutes between 6AM and 12AM. The average speeds early morning and late at night are generally quite high while it drops during the day and in the evening. This can be attributed to vehicular traffic at different times of the day - lower volumes of motor vehicles will result in higher average speeds.

\subsection*{Microblog Posts}
To testify the validation of our collected microblog data, we randomly choose 762 microblog posts from 20 April 2015. For each post, we categorize the words by their length and visualize them as a wordcloud.
As seen in Figure~\ref{fig:wordcloud}, the most frequent words in these posts are also the most commonly used Chinese words. This suggests that the data correctly reflects the social media in Beijing.

\subsection*{Advanced Analytics}
We comprehensively evaluate the proposed mixed dataset by referring to advanced data analytics conducted in other works. For example, PATE~\cite{zhao2022pate} uses the H4M dataset to assess the social and economic value of the real estate in Beijing. By estimating many factors of real estate prices, H4M boosts the accuracy of the price valuation model. Moreover,  Zhou \textit{et al.}~\cite{zhou2022traffic} leverages the dataset in predicting large-scale traffic congestion based on multi-modal fusion and representation mapping.

By respectively using a single factor and fusing diverse data, the aforementioned works perform ablation study on the dataset. Both studies conclude that the intrinsic heterogeneous, multi-source, multi-modal, multi-view and multi-distributional characteristics of H4M have practical applications and are also beneficial in improving modeling accuracy.

\begin{figure*}[h]
\centering
\includegraphics[width=\linewidth]{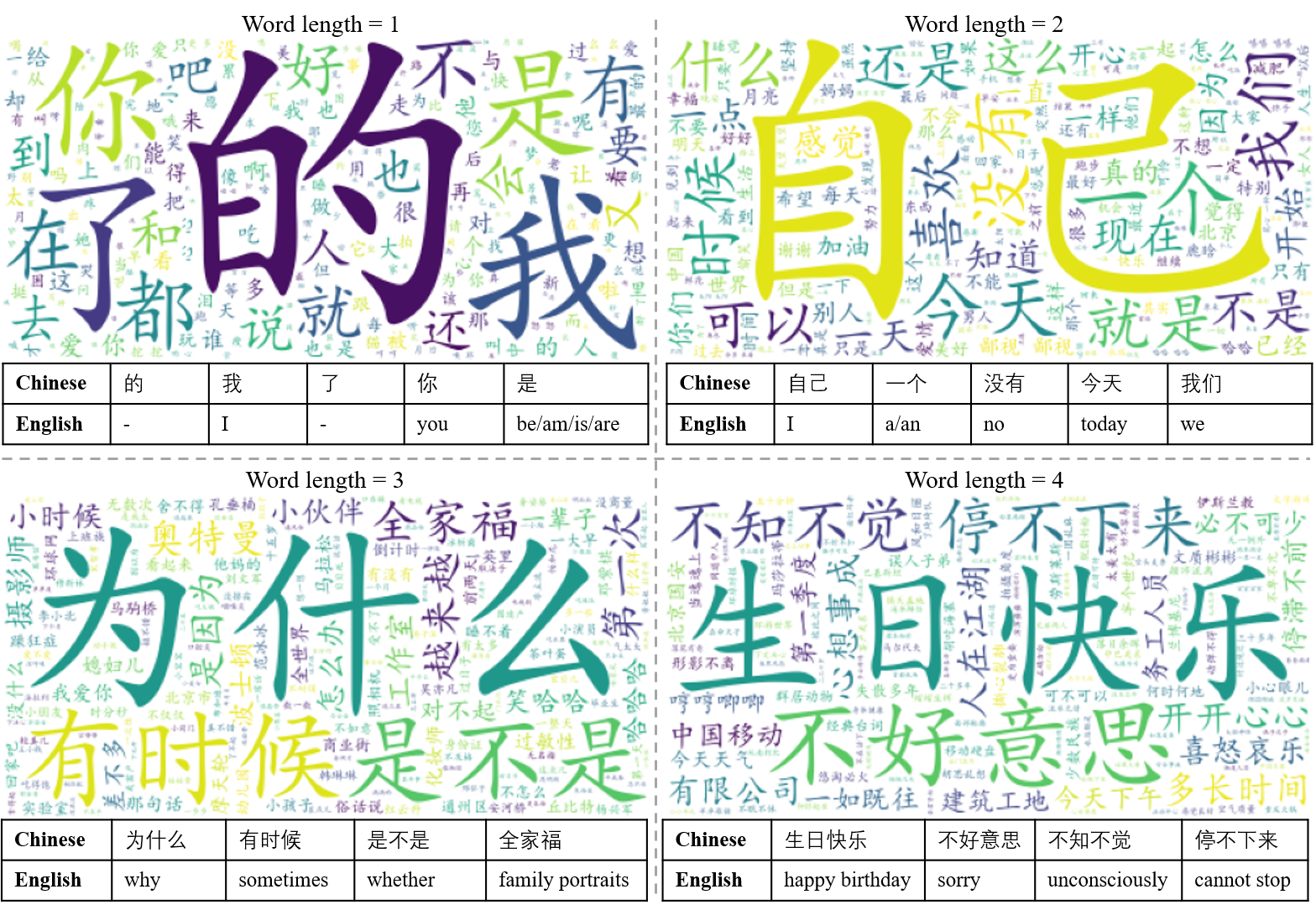}
\caption{Wordcloud visualization of words from 762 randomly chosen microblog posts on 20 April 2015. A Larger font size indicates higher frequency of the word. The table under each wordcloud contains the English translation of the most frequently used Chinese words.}
\label{fig:wordcloud}
\end{figure*}

\section*{Data Records}
In this section, we provide an overview of the data files and their formats. Given the heterogeneous nature of our dataset, it is imperative that we describe each different type of data.

\subsection*{Real Estate}
For housing analytics, we crawl through approximately $29000$ real estate data in Beijing - After cleaning, there are a total of $28645$ entries. Each entry has $21$ variables. The variables are defined below:

\begin{itemize}
    \item Id: the identity number of the points of interest.
    \item TotalPrice: selling price of the house in thousand Renminbi (RMB). RMB is the legal currency of China.
    \item UnitPrice: price  per square meter of the house in RMB.
    \item Year: building year.
    \item RoomNum: the number of bedrooms in the house.
    \item HallNum: the number of living and dining rooms in the house.
    \item KitchenNum: the number of kitchens in the house.
    \item BathNum: the number of bathrooms in the house.
    \item Floor: floor on which the house is located. Possible values are: low, middle, high.
    \item Orientation: the orientation of the house.
    \item Decoration: type of the house decoration.
    \item BuildingArea: the building area of the house.
    \item InsideArea: the inside area of the house.
    \item Heating: heating mode of the house.
    \item Elevator: whether there is an elevator in the building. The value is $1$ if there exists an elevator along with the house, $0$ otherwise.
    \item Layout: the type, area, orientation and window existence of each division in the house. This variable consists of $4$ sub-variables:
        \begin{itemize}
            \item Type: type of the house division.
            \item Area: area of the house division.
            \item Orient: orientation of the house division.
            \item Window: whether there exists a window in the house division. The value is $1$ if there exists an elevator along with the house, $0$ otherwise.
        \end{itemize}
    \item Lat: latitude of the house.
    \item Lng: longitude of the house.
\end{itemize}

\subsection*{Points of Interest}
In this paper, we crawl through $552,024$ pieces of Beijing points of interest data, and there are $497,256$ pieces varying in $18$ types after cleaning. Each data has $4$ variables. The variables are defined below:
\begin{itemize}
    \item Id: the identity number of the points of interest.
    \item Lat: the latitude of the points of interest.
    \item Lng: the longitude of the points of interest.
    \item Type: the type of the points of interest. Possible values are: office buildings, automobile repair, automobile services, automobile sales, public transport, retail goods, retail services, financial services, healthcare, serviced apartments, restaurants, education, public sports and leisure facilities, government buildings, tourist attractions, hotel services, traffic facilities and public facilities.
\end{itemize}

\subsection*{Traffic}
Traffic data is stored as a Pickle file consisting of $250,000$ elements corresponding to different regions in Beijing and each element is a list of $216$ values. These values are daily traffic speeds in km/h recorded every $5$ minutes (between 6 \textit{a.m.} and \textit{12 p.m.}).

\subsection*{Microblog Posts}
There are $460$ text files containing over $100$ million of microblog posts in Beijing. Each file contains rows with daily posts between September 12, 2013 and April 20, 2015. Each row has space separated values that are as follows: 
\begin{itemize}
    \item The content of the microblog.
    \item The longitude of the microblog location.
    \item The latitude of the microblog location.
    \item The post date and time of the microblog.
\end{itemize}

\begin{CJK}{UTF8}{gbsn}
\begin{table*}[h]
\centering
\begin{tabular}{|l|l|l|l|l|l|l|l|l|}
\hline
Main text & Longitude & Latitude & Week & Month & Day & Time & Time zone & Year \\
\hline
下雨喽，好开心，好凉快\dots & 116.142647 &	39.729572 &	Fri & Sep & 13 & 21:32:14 & +0800 & 2013\\
\hline

\end{tabular}
\vspace{2mm}
\caption{\label{tab:micro_parts}Interpretation of a piece of microblog data, where the time zone ``+0800" means Greenwich Mean Time (GMT) plus eight. }
\end{table*}
\end{CJK}

\section*{Usage Notes}
\subsection*{House Data}
The house data is stored in a JSON file. For instance, reading a piece of house data from the JSON file will output:

\noindent\{\\
    \indent``Id": ``101087602731",\\
    \indent``TotalPrice": ``24660",\\
    \indent``UnitPrice": “82533”,\\
    \indent``Year": ``2010",\\
    \indent``RoomNum": ``4",\\
    \indent``HallNum": ``1",\\
    \indent``KitchenNum": ``1",\\
    \indent``BathNum": ``3",\\
    \indent``Floor": ``low",\\
    \indent``Orientation": Southeast,\\
    \indent``Decoration": ``Penthouse/Exquisite Decoration",\\
    \indent``BuildingArea": ``298.79",\\
    \indent``InsideArea": ``242.89",\\
    \indent``Heating": ``Self-heating",\\
    \indent``Elevator": ``1",\\
    \indent``Layout": \\
        \indent\{\\
            \indent\indent``Type": [``living room", ``bedroom A", ``bedroom B", ``bedroom C", ``bedroom D", ``kitchen", ``bathroom A", \dots],\\
            \indent\indent``Area": [68.71, 34.29, 24.12, 14.74, 15.43, 11.87, 10.1, \dots],\\
            \indent\indent``Orient": [``South East", ``South", ``East", ``East", ``South East", ``None", ``None", \dots]\\
            \indent\indent``Window": [1, 1, 1, 1, 1, 0, 0, \dots]\\
        \indent\}\\
    \indent``Lat": 40.006694137576851,\\
    \indent``Lng": 116.48668712633133\\
    \}

\subsection*{Points of Interest}
The points of interest data is stored in a JSON file. For instance, reading a piece of points of interest data from the JSON file will output:

\noindent\{\\
    \indent``Id": ``110107",\\
    \indent``Lat": 39.963204,\\
    \indent``Lng": 116.125696,\\
    \indent``Type": ``office buildings"\\
\}

\subsection*{Traffic Data}
The traffic data is stored as a Pickle file. For instance, reading a piece of microblog data from a text file will output:

[0.0, 0.0, 0.0, 10.0, 10.0, 10.0, 0.0, 0.0, 23.0, 23.0, 23.0, 0.0, 6.0, 6.0, 6.0, 28.0, 28.0, 28.0, 26.0, 26.0, 26.0, 6.0, 6.0, 6.0, 23.0, 23.0, 23.0, 30.0, 30.0, 30.0, 30.0, 34.0, 34.0, 34.0, 26.0, 47.0, 47.0, 47.0, 47.0, 13.0, 13.0, 13.0, 15.0, 15.0, 15.0, 38.0, 38.0, \dots]

\subsection*{Microblog Data}
The Microblog data is stored as text files. For instance, reading a piece of microblog data from a text file will output:

\begin{CJK}{UTF8}{gbsn}
下雨喽，好开心，好凉快。我在这里:http://t.cn/z8AUByp	116.142647	39.729572	Fri Sep 13 21:32:14 +0800 2013
\end{CJK}

For a better understanding, we divide this example into parts and interpret their meaning as Table~\ref{tab:micro_parts} shows.


{
\bibliographystyle{unsrt}
\bibliography{sample}

\begin{thebibliography}{10}

\bibitem{london}
Binbin Lu, Martin Charlton, Paul Harris, and A~Stewart Fotheringham.
\newblock Geographically weighted regression with a non-euclidean distance
  metric: a case study using hedonic house price data.
\newblock {\em International Journal of Geographical Information Science},
  28(4):660--681, 2014.

\bibitem{fairfax}
Byeonghwa Park and Jae~Kwon Bae.
\newblock Using machine learning algorithms for housing price prediction: The
  case of fairfax county, virginia housing data.
\newblock {\em Expert systems with applications}, 42(6):2928--2934, 2015.

\bibitem{ames}
Dean De~Cock.
\newblock Ames, iowa: Alternative to the boston housing data as an end of
  semester regression project.
\newblock {\em Journal of Statistics Education}, 19(3), 2011.

\bibitem{lagos}
Adedeji~O Afolabi, Rapheal~A Ojelabi, BA~Adewale, Adedotun Akinola, and Adesola
  Afolabi.
\newblock Statistical exploration of dataset examining key indicators
  influencing housing and urban infrastructure investments in megacities.
\newblock {\em Data in brief}, 18:1725--1733, 2018.

\bibitem{pems}
Chao Chen, Karl Petty, Alexander Skabardonis, Pravin Varaiya, and Zhanfeng Jia.
\newblock Freeway performance measurement system: mining loop detector data.
\newblock {\em Transportation Research Record}, 1748(1):96--102, 2001.

\bibitem{tian2018lstm}
Yan Tian, Kaili Zhang, Jianyuan Li, Xianxuan Lin, and Bailin Yang.
\newblock Lstm-based traffic flow prediction with missing data.
\newblock {\em Neurocomputing}, 318:297--305, 2018.

\bibitem{short_text}
Hao Wang, Zhengdong Lu, Hang Li, and Enhong Chen.
\newblock A dataset for research on short-text conversations.
\newblock In {\em Proceedings of the 2013 Conference on Empirical Methods in
  Natural Language Processing}, pages 935--945, 2013.

\bibitem{weibo_rank}
Qing Liao, Wei Wang, Yi~Han, and Qian Zhang.
\newblock Analyzing the influential people in sina weibo dataset.
\newblock In {\em 2013 IEEE Global Communications Conference (GLOBECOM)}, pages
  3066--3071. IEEE, 2013.

\bibitem{weibo_cov}
Yong Hu, Heyan Huang, Anfan Chen, and Xian-Ling Mao.
\newblock Weibo-cov: A large-scale covid-19 social media dataset from weibo.
\newblock {\em arXiv preprint arXiv:2005.09174}, 2020.

\bibitem{quercia2012tracking}
Daniele Quercia, Jonathan Ellis, Licia Capra, and Jon Crowcroft.
\newblock Tracking" gross community happiness" from tweets.
\newblock In {\em Proceedings of the ACM 2012 conference on computer supported
  cooperative work}, pages 965--968, 2012.

\bibitem{llorente2015social}
Alejandro Llorente, Manuel Garcia-Herranz, Manuel Cebrian, and Esteban Moro.
\newblock Social media fingerprints of unemployment.
\newblock {\em PloS one}, 10(5):e0128692, 2015.

\bibitem{milan}
Gianni Barlacchi, Marco De~Nadai, Roberto Larcher, Antonio Casella, Cristiana
  Chitic, Giovanni Torrisi, Fabrizio Antonelli, Alessandro Vespignani, Alex
  Pentland, and Bruno Lepri.
\newblock A multi-source dataset of urban life in the city of milan and the
  province of trentino.
\newblock {\em Scientific data}, 2(1):1--15, 2015.

\bibitem{zhao2022pate}
Yaping Zhao, Ramgopal Ravi, Shuhui Shi, Zhongrui Wang, Edmund~Y Lam, and
  Jichang Zhao.
\newblock Pate: Property, amenities, traffic and emotions coming together for
  real estate price prediction.
\newblock In {\em IEEE International Conference on Data Science and Advanced
  Analytics}. IEEE, 2022.

\bibitem{zhou2022traffic}
Bodong Zhou, Jiahui Liu, Songyi Cui, and Yaping Zhao.
\newblock Large-scale traffic congestion prediction based on multimodal fusion
  and representation mapping.
\newblock In {\em IEEE International Conference on Data Science and Advanced
  Analytics}. IEEE, 2022.

\end{thebibliography}
}

\end{document}